\documentclass[nofootinbib,12pt]{revtex4}
\usepackage{graphicx,amssymb,amsfonts,amsthm,amsmath,amssymb,amscd,amstext}
\def\be{\begin{equation}}
\def\ee{\end{equation}}

\def\sinh{\operatorname{sinh}}
\def\coth{\operatorname{coth}}
\def\log{\operatorname{log}}
\def\sin{\operatorname{sin}}

\def\arcsin{\operatorname{arcsin}}
\def\arctan{\operatorname{arctan}}
\def\Area{\operatorname{Area}}
\def\EllipticK{\operatorname{EllipticK}}

\begin{document}

\title{Constructing Space From Entanglement Entropy}
\author{Michael Spillane\footnote{email: michael.spillane@stonybrook.edu}}
\affiliation{\it C.~N.~Yang Institute for Theoretical Physics \\ Stony Brook University, Stony Brook, NY  11794}

\begin{abstract}
We explicitly reconstruct the metric of a gravity dual to field theories using known entanglement entropies using the Ryu-Takayanagi formula \cite{Ryu:2006bv}.  We use for examples CFT's in $d = 1$, 2 and 3 as well as CFT on a circle of length $L$ and a thermal CFT at temperature $\beta^{-1}$.  We also give the first several coefficients in the Taylor series of the metric for a general entanglement entropy in 1+1 dimensions as well as some examples (Appendix B).  The beginnings of a dictionary between the dual theories appears naturally and does not need to be inserted by hand. For example, the dictionary entries $c=3R/2G_N$ for 1+1 dimensional CFT and $N^2 = \pi R^3/2G_N$ for $\mathcal{N}=4$ SYM in 3+1 dimensions are forced upon us.  After uploading this paper I was made aware of \cite{Bilson} which solves the same problem in a similar way.
\end{abstract}

\maketitle 

\section{Introduction}

The AdS/CFT correspondence, more generally Gauge/Gravity Duality, has presented short cuts for computing quantities that were difficult previously by considering a dual problem.  Such examples include thermodynamic properties of strongly coupled plasmas \cite{Kovtun} and energy loss of heavy quarks in such plasmas \cite{Herzog06}.  Recently a program was proposed for calculating the entanglement entropy of a given field theory using the gravity dual of the field theory  \cite{Ryu:2006bv},\cite{Lewkowycz}.    Entanglement entropy is a potentially useful order parameter to calculate for the study of topological phases of matter.  It is also of interest in RG flow \cite{Casini04},\cite{Myers12}. However, entanglement entropy is a notoriously difficult quantity to calculate in a general field theory context and so this proposal provided a welcome tool for analyzing field theories with known gravity duals.

Since its discovery there has been interest in how holography emerges from field theory and there is some reason to believe that entanglement entropy provides a nice way of investigating this possibility \cite{swingle}.  Our paper makes an effort in this direction to explicitly show how the metric can be calculated from a known entanglement entropy.  While the metric is only the tip of the iceberg for the gravity dual it is a reasonable place to start.

Our starting point for recovering the metric of a dual field theory is the Ryu-Takayanagi formula\cite{Ryu:2006bv}
\begin{align}
S_A = \frac{\Area(\gamma_A)}{4 G_N}.
\end{align}
Where $\gamma_A$ is the minimal surface which corresponds to the region $A$ on the conformal boundary, $S_A$ is the entanglement entropy between $A$ and $\overline{A}$ and $G_N$ is the Newton's constant in the gravity theory.
We start by making the following ansatz for the metric
\begin{align}
\label{metric}ds^2=\frac{R^2}{z^2}\left(g(z)^2 dz^2+dx^2+\sum_{i=2}^{d-1}dx_i^2\right).
\end{align}
Where $R$ is a constant which in pure AdS corresponds to the radius.
We will only consider the strip geometry, namely that $x_1 = x$ can vary while all other $x_i$ are held fixed.  Without loss of generality we can then restrict ourselves to the case where $g(0)=1$ because any overall constant can be absorbed into $R$. As we will see requiring $g(0)=1$ will allow us to generate the start of a dictionary between the field theory and its gravity dual.  This ansatz is the most general time independent, translationally invariant metric, up to coordinate redefinitions.  We next  need to find the equations for the minimal surface from which we can determine the entanglement entropy. 

\begin{align}
\Area(\gamma_A) \propto \int \frac{dz}{z^d}\sqrt{g(z)^2+x'(z)^2}.
\end{align}

Using the Euler-Lagrange equation we then obtain

\begin{align}
x'(z)^2= \frac{g(z)^2 z^{2d}}{z_*^{2d}-z^{2d}},
\end{align}
where $z_*$ is maximum $z$ which the minimal surface reaches in the bulk.  We can then take this solution and obtain two coupled integral equations which must be solved in order to determine the metric.  The equations read
\begin{align}
\label{length} \ell/2 &= \int_0^{z_*}dz x'(z) = \int_{0}^{z_*} dz \frac{g(z) z^d}{\sqrt{z_*^{2d}-z^{2d}}}, \\
\label{entropy}S_A &= \frac{R^dL^{d-1}}{2G_N}\int_\epsilon^{z_*}\frac{dz}{z^d}g(z)\frac{z_*^d}{\sqrt{z_*^{2d}-z^{2d}}}.
\end{align}
Where $L$ is the length of the strip in the $x_i$ directions.
The first equation I will call the length equation and the second the entropy equation.  The integrand in the integral equation is positive; therefore $z_*$ is in one to one correspondence with $\ell$ and so in principle we can convert between the two as needed.

Comparing the two sides (field theory and AdS) will then give a relationship between the UV cutoffs and also a relationship between the field theory variables and $R$ and $G_N$.  Considering (\ref{entropy}) for $g_0(z)=1$ we can get the relationship between the field theory and the gravity by looking at the first term in a Taylor series of $S_A$.

After uploading this paper I was made aware of \cite{Bilson} which solves the same problem in a similar way.

\section{1+1 Dimensions}

Let us  now consider the simplest case of $d=1$.   It is helpful to take the entropy equation (\ref{entropy})  and differentiate with respect to $z_*$ which will remove the UV cutoff. In doing so we also remove terms which are independent of the length of the interval. However, such terms are usually non universal as they can be altered by a redefinition of the UV cutoff.  Taking the derivative results in
\begin{align}
(\partial_{z_*}\ell) (\partial_\ell S_A) = \frac{R}{2G_N}\left(\int_0^1dy\frac{g'(z_* y)}{\sqrt{1-y^2}}+\frac{1}{z_*}\right).
\end{align}
where we have used $g(0)=1$ as well as taking the limit $\epsilon \rightarrow 0$ .  Rearranging yields
\begin{align}
\label{A}\int_0^{z_*}dz \frac{g'(z)}{\sqrt{z_*^2-z^2}}=\frac{2G_N}{R}(\partial_{z_*}\ell)(\partial_\ell S_A)-\frac{1}{z_*}.
\end{align}

For general $S_A$ the coupled equations probably cannot be solved exactly.  However, we can assume that $g(z)$ is given by a Taylor series in $z$ and solve for the elements by requiring that $g$ satisfies both equations simultaneously.  In doing so there is a possibility that solutions which are non-analytic functions at $z=0$ are overlooked. We use equation (\ref{length}) to determine $\ell$ in terms of $z_*$.  We can then solve equation (\ref{entropy}) for $g(z)$ and then require that the test $g$ we used to calculate $\ell$ is the same as this solution. 

To invert this equation we will use the following identity \footnote{When using a Taylor series for $g$ there is actually no need to invert the equations. Rather you can calculate the Taylor series on both sides and match coefficients.  However, this inversion allows one to carry out an iterative procedure which may be useful for numerically solving these equations (see figure \ref{convergenceplot})}
\begin{align}
\label{identity}\int_t^y dx\frac{2x}{\sqrt{(x^2-t^2)(y^2-x^2)}} = \pi
\end{align}
First we multiply both sides of (\ref{A}) by $2z_*/\sqrt{z_*^2-x^2}$ and integrate from 0 to $y$ with respect to $x$.
This yields, after switching the order of integration on the double integral, 
\begin{align}
\int_0^y dz \int_z^{y}dz_* \frac{2z_*g'(z)}{\sqrt{(z_*^2-z^2)(y^2-z_*^2)}}=\left(\frac{2G_N}{R}\int_0^ydz_*\frac{2z_*}{\sqrt{y^2-z_*^2}}(\partial_{z_*}\ell)(\partial_\ell S_A)\right)-\pi
\end{align}
Using equation (\ref{identity}) and the fundamental theorem of calculus yields
\begin{align}
\label{iteration}g(y)=\frac{1}{\pi}\frac{2G_N}{R}\int_0^ydz_*\frac{2z_*}{\sqrt{y^2-z_*^2}}(\partial_{z_*}\ell)(\partial_\ell S_A)
\end{align}
Observe that this implies that if $S_A$ is an odd function of $\ell$ then $g(y)$ is an even function of $y$. The next 3 examples have $S_A$ odd so the resulting $g$ must be even which slightly shortens our calculations.  The three examples we will consider are a CFT, a finite temperature CFT and a CFT on a circle of length $L$.  These three have well known holographic dual metrics and are therefore good choices to test the system developed here.  In appendix B some other examples are worked out exactly.  One is a candidate for a field theory with RG flow and the other are simple examples which can be solved exactly.

\subsection{Conformal Field Theory}
The simplest case to consider is that of a CFT.  For a single interval the result is known and takes the form \cite{Holzhey}

\begin{align}
S_A=\frac{c}{3}\log[\ell/\epsilon].
\end{align}

We are interested in showing that $g(z)=1$ is the unique solution to the above equations (at least among analytic functions).  Using the fact that $\partial_\ell S_A$ is odd we first take the following Taylor expansion of  $g(z)$
\begin{align}
g_0(z) =1+ \sum_{n=1}^\infty a_{2n} z^{2n}.
\end{align}
We can now plug $g_0(z)$ into the length equation to obtain
\begin{align}
\ell &= 2x+\sqrt{\pi}\sum_{n=1}^\infty a_{2n} x^{2n+1}\frac{\Gamma(n+1)}{\Gamma(\frac{2n+3}{2})}.
\end{align}
Next we plug this result into (\ref{iteration})

\begin{align}
g(z) &= \frac{1}{\pi}\int_0^z dx \frac{2x}{\sqrt{z^2-x^2}}\frac{\ell'(x)}{\ell(x)}
\end{align}
We can then perform a Taylor expansion of $1/\ell$ to obtain
\begin{align}
\label{B}g(z) &=\frac{2  G_N c}{3R}\left(1+\frac{2 a_2}{3}z^2+\left(\frac{4 a_4}{5}-\frac{a_2^2}{3}\right)z^4+...\right)
\end{align}
We now require that the $g(z)$ we calculated is equal to $g_0(z)$, i.e. we are at a``fixed point".  First we recover the well known result that $2G_Nc/3R=1$  \cite{Brown}.   Looking at (\ref{B}) further we observe that $a_n$ always first appears at order $z^n$ in the form $n a_n/(n+1)$.  We can therefore prove by induction that $a_n =0$ for all $n$.\footnote{It is clear from (\ref{B}) that $a_2=0$. If we then assume that $a_n = 0$ for all $n<m$ then the Taylor expansion of $g(z)$= 1+ $m a_mz^m/(m+1)$+... This then implies that $a_m=0$, so, by induction $a_n = 0$ for all $n$.}

\subsection{Finite Temperature}

We have shown for the simplest case that this method of calculating the metric works, but what about for more complicated situations? We can repeat the above process for a finite temperature, $\beta^{-1}$, conformal field theory.  We recall that the entanglement entropy is given by \cite{Korepin04}

\begin{align}
S_A= \frac{c}{3}\log\left(\frac{\beta}{\pi \epsilon}\sinh(\ell \pi/\beta)\right) \\
\label{derivative}\partial_\ell S_A = \frac{c}{3}\frac{\pi}{\beta}\coth(\ell \pi/\beta).
\end{align}
The approach taken in the section on conformal field theories (at zero temperature) results in extremely messy equations.  However, as noted above that $a_n$ first appears at order $z^n$ is a general feature (consider a Taylor expansion of $S_A$) and so we can determine $a_2$ first then $a_3$ and so on.
To do this consider the following $g$. (I have set $\beta = \pi$ to make the equations simpler. The $\beta$ dependence can be recovered by the transformation $z\rightarrow z\pi/\beta$.)
\begin{align}
\label{ansatz}g_0(z)=1+a_2z^2.
\end{align}
Then plugging (\ref{ansatz}) into the length equations (\ref{length}) produces
\begin{align}
\ell(z) = 2z +4 a_2 z^3/3
\end{align}
We can then expand (\ref{derivative}) in a Taylor series in $z$ and plug that into (\ref{iteration}) to get \footnote{As with the conformal field theory case to get $g(0)$ =1, we require $c = 3R/2G_N$.}
\begin{align}
g(z) = 1+\frac{2}{3}(1+a_2)z^2+O(z^4).
\end{align}

Requiring that $g$ is fixed under iteration then gives $a_2=2.$\footnote{One can also solve the resulting recurrence relation to see that $a_{2,\infty} = 2.$}  We now obtain $a_4$ by using 
\begin{align}
g_0(z)=1+2z^2+a_4z^4.
\end{align}
Repeating the above process yields
\begin{align}
g_1(z) = 1+2z^2+(6/5+4a_4/5)z^4+O(z^6).
\end{align}
So $a_4 = 6$.
Repeating this yields the following $a_{2n+1}=0, a_4= 6,a_6=20,a_8=70$.  One can recognize these numbers as the first few coefficients in the Taylor expansion of 
\begin{align}
\label{C}g(z)=\frac{1}{\sqrt{1-4z^2}}
\end{align}

 Plugging (\ref{C}) into our ansatz for the metric (\ref{metric}) yields the black brane metric for the BTZ black hole  as is expected \cite{BTZ}.  The coordinate transformation is given by $z=R/r$ and $x\rightarrow\phi$. Note the $dt$ portion of the metric has been left out. Which results in 

 \begin{align}
ds^2 = \frac{R^2dr^2}{r^2-(2\pi R/\beta)^2}+r^2d\phi^2
\end{align}

\subsection{Spatial Circle}

The entanglement entropy for a zero temperature CFT on a circle is given by \cite{Calabrese04}

\begin{align}
S_A = \frac{c}{3}\log\left(\frac{L}{\pi\epsilon} \sin(\pi \ell/L)\right)
\end{align}
Repeating the above approach we can calculate the first several terms in the Taylor expansion to obtain $a_{2n+1}=0, a_2= -2 (\pi/L)^2, a_4= 6(\pi/L)^4,a_6=-20(\pi/L)^6,a_8=70(\pi/L)^8.$  We can recognize these as the coefficients of the Taylor expansion of
\begin{align}
g(z) = \frac{1}{\sqrt{1+(2\pi z/L)^2}}.
\end{align}

Plugging (\ref{C}) into our ansatz for the metric (\ref{metric}) yields 
\begin{align}
ds^2=\frac{R^2}{z^2}\left(\frac{dz^2}{1+4\pi^2z^2/L^2}+dx^2\right).
\end{align}
From this metric it is not obvious how it is related to a spatial circle.  However via the coordinate transformation $\omega = z/\sqrt{1+(2\pi z/L)^2}$ we can cast this metric in a more familiar form, of the AdS soliton \cite{soliton},
\begin{align}
ds^2=\frac{R^2}{\omega^2}\left(\frac{d\omega^2}{1-4\pi^2\omega^2/L^2}-(1-4\pi^2\omega^2/L^2)dx^2\right).
\end{align}

\subsection{General EE}
One could repeat this process for a general Taylor expansion of the entanglement entropy.  From doing so, for $\partial_\ell S_A = \alpha/\ell +\sum_{n=0}^\infty b_n\ell^n$, the following results come out:
\begin{enumerate}
  \item  Any terms more divergent than $\ell^{-1}$ result in a divergent integral (\ref{iteration}) and are therefore excluded.  The $1/\ell$ term also sets  the relationship between the field theory content and the gravity to be $2G_N\alpha/R=1$.  If there is no such divergent term ($1/\ell$) in the $\ell \rightarrow 0$ limit, the holographic geometry is not asymptotically AdS. 
  \item For $g(z)$ = 1+ $\sum_{n=1}^\infty a_nz^n$ the coefficients are presented in the following table
  \begin{center}
  \begin{tabular}{ | l | p{15cm} | }
    \hline
    $n$& $a_n$ \\ \hline
    $1$ & 8$b_0$/$\pi$  \\ \hline
    $2$ & 6($b_0^2$+$b_1$)  \\ \hline
    $3$ & 128($b_0^3$+3$b_0b_1$+$b_2$)/3$\pi$  \\ \hline
    $4$ & 30($b_0^4+6b_0^2b_1+2b_1^2+4b_0b_2+b_3)$  \\ \hline
    $5$ & 1024($b_0^5+10b_0^3b_1+10b_0^2b_2+5b_1b_2+5b_0(2b_1^2+b_3)+b_4)/5\pi$  \\ \hline
    $6$ & 140($b_0^6+15b_0^4b_1+5b_1^3+20b_0^3b_2+3b_2^2+ 6b_1b_3+15b_0^2(2b_1^2+b_3)+ 6b_0(5b_1b_2+b_4)+b_5))$  \\ \hline
    $7$ & 32768($b_0^7+21b_0^5b_1+35b_0^4b_2+35b_0^3(2b_1^2+b_3)+7b_2(3b_1^2+b_3)+7b_1b_4 +21b_0^2(5b_1b_2+b_4)+7b_0(5b_1^3+3b_2^2+6b_1b_3+b_5)+b_6)/35\pi$  \\ \hline
  \end{tabular}
  \end{center}
   Observe that if $\partial_\ell S_A$ is an odd function of $\ell$ then $g$ is an even function as mentioned previously.
  \end{enumerate}

\subsection{Recursion}
While the above approach finds the metric when the functional form of the entanglement entropy is known, one might wish to find metrics for cases where the EE is generated numerically.  One approach is to put in a test function for $g(z)$ and see if iterating the above procedure- plug in the $g(z)$ recovered from (\ref{iteration}) back in as the new $g(z)$- will eventually converge to the solution. 

We first consider the pure CFT by solving the recursion relations that result from (\ref{B}).  Let $a_{n,m}$ be the $m^{th}$ iteration of $a_n$ so $g_m(z) = 1+\sum_{n=1}^\infty a_{n,m}z^n$.

Then
\begin{align}
a_{2,m}=2a_{2,m-1}/3 
\end{align}
The solution is easy to find as
\begin{align}
a_{2,m}=a_{2,1}(2/3)^{m-1}.
\end{align}
So in the $\lim_{n\to\infty}a_{2,n}=0$ and the convergence is exponential.
The next relation is for $a_4$.  Which has the relation
\begin{align}
a_{4,m}=4 a_{4,m-1}/5-a^2_{2,m-1}/3.
\end{align}
The solution is given by
\begin{align}
a_{4,m}=\left(\frac{4}{5}\right)^{m-1}a_{4,1}+\left(5\left(\frac{3}{4}\right)^3(2/3)^{2m}-\frac{3}{4}(4/5)^{m-2}\right)a_{2,1}^2
\end{align}
Which also converges exponentially to zero as $m\to \infty$.  This exponential convergence is true generically for the convergence for all coefficients (See Appendix A).  The convergence, however, is slower for higher order terms and goes as $(n/n+1)^m$.  

\begin{figure}
 \begin{center}
a) \includegraphics[width=2.9in]{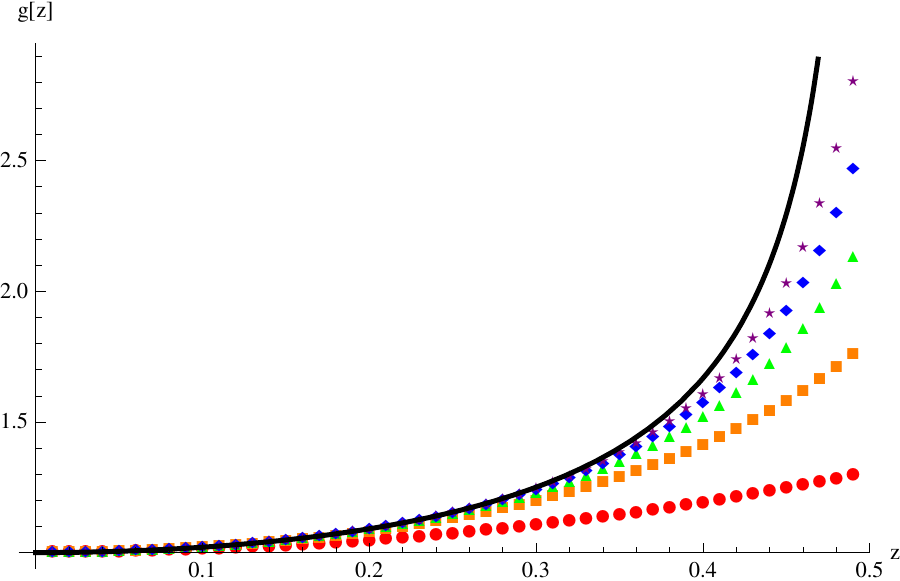} 
b) \includegraphics[width=2.9in]{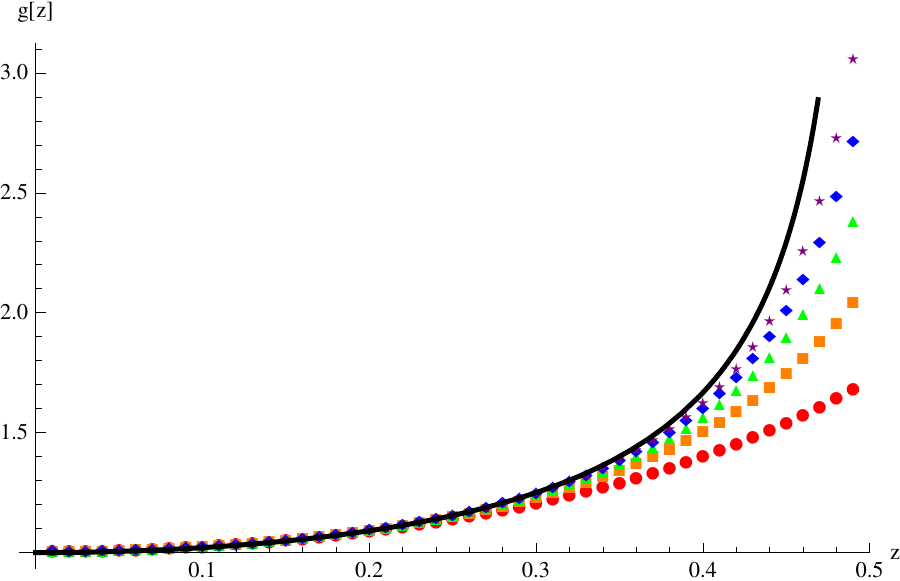} 
 \end{center}
 \caption{
a) This figure shows the convergence of an iterated test function to the solution.  The test function was taken to be $g_0 = 1$.  b) For this figure the initial test function was $g_0 = 1+2z^2$. In both cases $g_2$, $g_6$, $g_{10}$, $g_{14}$ and $g_{18}$ are plotted, where those closer to the solution correspond to high iterations.  As expected the convergence is faster for a better test function.
 \label{convergenceplot}
  }
 \end{figure}

We can repeat this process for the thermal CFT.  As seen in the conformal field theory for a general test function $g_0$ converges to this answer under iteration (see Appendix A).  The convergence of two difference test functions was done numerically and is shown in Figure \ref{convergenceplot}.

 \begin{figure}[htbp]
 \begin{center}
\includegraphics[width=3.5in]{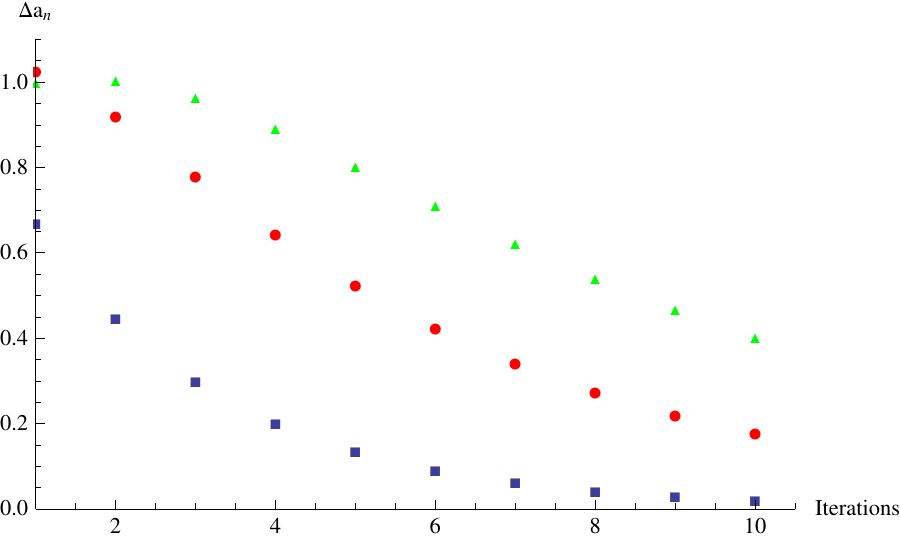}  
 \end{center}
 \caption{
a) This figure shows the convergence of the coefficients of $g$ to the solution.  Here $\Delta a_n = (a_\infty-a_n)/a_\infty.$ This figure shows that convergence is slower for higher order terms in the Taylor series.  The blue squares are the $z^2$ coefficient, red circles are $z^4$ coefficients and green triangles are $z^6$ coefficients.
  }
 \end{figure}
 
In fact for any test function and any $\partial_\ell S_A$ which is a analytic function of $\ell$ this iterative process converges to the unique, analytic solution (Appendix A).

\newpage
\section{Time Component of Metric}
For the above calculations we took a constant time slice of the metric and performed the calculation.  We would like to be able to reconstruct the time component of the metric.  Perhaps using some additional knowledge about the field theory and the how it is calculated in the dual theory one could reconstruct the time component.  However, in the three explicit cases discussed the time component can be reproduced by assuming that the metric is diagonal and satisfies the vacuum Einstein Equation with cosmological constant -1/$R^2$.

\subsection{Conformal Field Theory}

Above we calculated that for a constant time slice the spatial metric was 
\begin{align}
ds^2=\frac{R^2}{z^2}(dz^2+dx^2).
\end{align}
We make the following ansatz for the metric
\begin{align}
ds^2=\frac{R^2}{z^2}(f(z) dt^2+dz^2+dx^2).
\end{align}
One can then calculate the Ricci tensor and scalar curvature and plug them into the Einstein equation for vacuum.  This results in the following two equations

\begin{align}
&f'(z)=0 \\
&z f'(z)^2+2f(z)(f'(z)-zf''(z))=0
\end{align}

It is easy to see that these equations are redundant and admit the solution $f(z) = c$ for some constant $c$. This is the Poincar\'{e} patch of AdS$_3$.

\subsection{Thermal CFT}
Above we calculated that for a constant time slice the spatial metric for a thermal CFT was 
\begin{align}
ds^2=\frac{R^2}{z^2}\left(\frac{1}{1-4\pi^2z^2/\beta^2}dz^2+dx^2\right).
\end{align}
We make the following ansatz for the metric
\begin{align}
ds^2=\frac{R^2}{z^2}\left(f(z) dt^2+\frac{1}{1-4\pi^2z^2/\beta^2}dz^2+dx^2\right).
\end{align}
Requiring that the vacuum Einstein equation is satisfied gives the two following equations
\begin{align}
&8\pi^2 z f(z)+(\beta^2-4\pi^2z^2)f'(z)=0 \\
&z(\beta^2-4\pi^2z^2)f'(z)^2-2f(z)(\beta^2f'(z)+z(4\pi^2z^2-\beta^2)f''(z)=0.
\end{align}
As before these equations are redundant.  They admit the solution 
\begin{align}
f(z)=c(1-4\pi^2z^2/\beta^2).
\end{align}
Via the coordinate redefinition $z = R/r$ gives the BTZ black hole \cite{BTZ}.
\subsection{Spatial Circle}
Above we calculated that for a constant time slice the spatial metric for a spatial circle was 
\begin{align}
ds^2=\frac{R^2}{z^2}\left(\frac{1}{1+4\pi^2z^2/L^2}dz^2+dx^2\right).
\end{align}
We make the following ansatz for the metric
\begin{align}
ds^2=\frac{R^2}{z^2}\left(f(z) dt^2+\frac{1}{1+4\pi^2z^2/L^2}dz^2+dx^2\right).
\end{align}
For the spatial circle we obtain the two following equations
\begin{align}
&8\pi^2 z f(z)-(L^2+4\pi^2z^2)f'(z)=0 \\
&z(L^2+4\pi^2z^2)f'(z)^2+f(z)(-2L^2f'(z)+2z(L^2+4\pi^2z^2)f''(z)=0.
\end{align}

As before these equations are redundant.  They admit the solution 
\begin{align}
f(z)=c(1+4\pi^2z^2/L^2)
\end{align}

As noted earlier this particular form of the metric is not particularly illuminating.  However, using the coordinate transformation $\omega = z/\sqrt{1+4\pi^2z^2/L^2}$ one obtains the AdS soliton \cite{soliton} in 2+1 dimensions,
\begin{align}
ds^2=\frac{R^2}{\omega^2}\left(-dt^2+\frac{d\omega^2}{1-4\pi^2\omega^2/L^2}-(1-4\pi^2\omega^2/L^2)dx^2\right).
\end{align}

\section{Higher Dimensions}
For more than 2 dimensions the equations become slightly messier. The area law predicts that (\ref{entropy}) diverges faster in the $\epsilon\rightarrow 0$ limit and so more derivatives are required to make the integral converge.  For these higher dimensional cases there are fewer known examples to test our procedure on so we will consider only zero temperature CFT's. 

\subsection{d=2 CFT}
For this case the entanglement entropy should take the form based upon the area law \cite{Holzhey}  
\begin{align}
S_A = c_1 \frac{L}{\epsilon}+c_2\frac{L}{\ell}
\end{align}
where L is the length of the surface in the $x_1$ direction and for some constants $c_1$ and $c_2$.  We notice that as with the 1+1 dimensional case an $\ell$ derivative kills the dependence on the cutoff.  So rearranging (\ref{entropy}) yields
\begin{align}
z_*S_A = \frac{R^2L}{2G_N}\int_{\epsilon/z_*}^1dy\frac{g(z_*y)}{y^2\sqrt{1-y^4}}.
\end{align}
Taking a derivative of each side gives
\begin{align}
\label{l derivative}\left(c_1 \frac{L}{\epsilon}+c_2\frac{L}{\ell}\right)\partial_\ell z_*-z_*c_2\frac{L}{\ell^2} = \frac{R^2L}{2G_N}\left(\int_{0}^{z_*}dz\frac{g'(z)z_*^2\partial_\ell z_*}{z\sqrt{z_*^4-z^4}}+\frac{g(0)\partial_\ell z_*}{\epsilon}\right)
\end{align}
This equation has parts dependent on the cutoff and independent of the cutoff.  We therefore choose the parts dependent on the cutoff and the not depending on the cutoff must be equal separately.  To make sure that the epsilon dependent parts cancel we get a relationship between the UV cutoff's of the two series.  To get a relationship between the field theory and gravity we need to compute (\ref{entropy}) for $g_0(z) =1$.
The result is
\begin{align}
& \frac{c_1}{\epsilon_{FT}}=\frac{R^2}{2G_N\epsilon_G} \\
&\label{c2}c_2 =-\frac{\pi R^2 \Gamma(3/4)^2}{\Gamma(1/4)^2G_N}
\end{align}
We would now like to invert this integral equation as we did before. For this we will need the following generalization of (\ref{identity}) for all $\alpha\neq0$.

\begin{align}
\label{AA}\int_t^ydx\frac{\alpha x^{(\alpha-1)}}{\sqrt{(y^\alpha-x^\alpha)(x^\alpha-t^\alpha)}}=\pi
\end{align}

Using (\ref{AA}) leads to
\begin{align}
\frac{\pi R^2 L}{G_N} \int_0^yg'(z)/z = \int_0^y\frac{4z_*}{\sqrt{y^4-z^4}}\left(c_2\frac{L}{\ell}-(\partial_{z_*}\ell)z_*c_2\frac{L}{\ell^2} \right)
\end{align}
One could complete the inversion process by taking a derivative with respect to $y$, multiplying by $y$, and then integrating. \footnote{This may be necessary for the iterative process described above and for numerical approaches.}  However, because we are going use a Taylor series we can now just plug in a Taylor expansion ansatz for $g$ and compare the results on both sides.  
We use the ansatz
\begin{align}
g(z) = 1+\sum_{n=2}^\infty a_n z^n.
\end{align}
Now using (\ref{c2}) for the $n^{th}$ term in the Taylor series, comparing the left and right hand sides gives (after using the solution for lower order coefficients, i.e. by induction on $n$)
\begin{align}
\frac{n}{n-1} a_n = \frac{n}{n+1}a_n
\end{align}
This means than $a_n =0$ for all $n\geq 2$ and $a_1=0$ because $S_A$ is an odd function. Therefore $g(z)= 1$.

\subsection{$\mathcal{N}$=4 SYM}
For higher dimensions one needs to take more derivatives with respect to $z_*$ as we did in (\ref{l derivative}) to make the integral convergent in the limit $\epsilon \rightarrow 0$. To this end we obtain
\begin{align}
\label{SYM}\partial_{z_*}^3(z_*^2S_A) = \frac{R^3L^2}{4G_N}\left(\frac{2g''(0)}{z_*}+\int_0^{z_*}dz\frac{2z_*^2g'''(z)}{\sqrt{z_*^6-z^6}}\right)
\end{align}
For $\mathcal{N}$ = 4 SYM we again use the result obtained in  \cite{Ryu:2006bv}
\begin{align}
S_A = \frac{N^2L^2}{2\pi\epsilon^2}-2\sqrt{\pi}\left(\frac{\Gamma(2/3)}{\Gamma(1/6)}\right)^3\frac{N^2L^2}{\ell^2}
\end{align}
The relationship between the field theory and gravity is given by
\begin{align}
\label{epsilon}\frac{N^2}{\pi \epsilon_{FT}^2} &= \frac{R^3}{2G_N\epsilon_G^2}\\
\label{N2}N^2 &= \frac{\pi R^3}{2G_N}
\end{align}
Solving these gives us the well known relationship for AdS/CFT and says that the gravity UV cutoff is the same as the CFT cutoff.

We can now use the now familiar Taylor expansion for $g(z)$ in (\ref{SYM}) and equating the two sides and using the dictionary  we found above, (\ref{epsilon}) and (\ref{N2}), to obtain

\begin{align}
a_2 &= 0\\
a_3 &= 0\\
 a_n&=\frac{n-2}{n+1} a_n \quad \text{for} \quad n\geq 4
\end{align}
So $g(z)$=1 is the unique solution.
It seems reasonable to assume that for any $d$ the condition will be $a_n = (n+1-d)/(n+1) a_n$.

\section{Conclusion}

We have seen that for all analytic forms of the entanglement entropy a unique analytic metric can be found which reproduces the entanglement entropy via the Ryu-Takayanagi formula.  One consequence of this is that in the strict gravity limit there is no new information in calculating two interval entanglement entropies.  This is of course not true in general; for example in a 2d CFT the answer is known to depend on all the operator content of the theory.  One could also look at numerical solutions to the system of integral equations particularly in looking for metrics which would match field theories for which the holographic dual is not known, for example SU($N$) gauge theories.

Future investigations along this line could include looking at different expansions of $\partial_\ell S_A$ including log terms to $\partial_\ell S_A$ which would need to be compensated by log terms in $g(z)$.  If there are no log terms in $\partial_\ell S_A$, there are no log terms in $g(z)$.  Other expansions may include a large length rather than small length expansion.  There might also be other ans\"{a}tze for the metric which would be better suited for looking at such alternate expansions.  There is a proposal for a time dependent formulation of holographic entanglement entropy \cite{Hubeny}.  One could then investigate how to generalize the method presented in this paper to that proposal.

\section{Acknowledgements}
I would like to acknowledge Christopher Herzog for his guidance and suggestions on the scope and direction during this project and Kristan Jensen for his comments and suggestions on this paper.  This work was supported in part by the National Science Foundation under Grants No. PHY-0844827 and PHY-1316617.

\section{Appendix A}
We would like to prove for the $d=1$ case that for any $\partial_\ell S_A$ which has a Taylor series expansion, that the iterative process converges exponentially to the correct solution for any initial test function.  By exponential convergence I mean that each element in the Taylor expansion converges exponentially fast.  As indicated in the text the proof is by induction so let us first consider the order $z$ term in the Taylor series.  We plug the following ansatz into (\ref{length})
\begin{align}
g_0(z)=1+a_1z^1 + O(z^2).
\end{align}
We then use (\ref{iteration}) to obtain 
\begin{align}
g_1(z)=1+(a_1/2+4b_0/\pi)z+O(z^2).
\end{align}
It is easy to see then that $a_{1,m}$ converges exponentially to 8$b_0/\pi$.  Assume that $a_{n,m}$ converges exponentially for all $n<N-1$.  Then the recursion relationship for $a_{N,m}$ is given by
\begin{align}
\label{recursion}a_{N,m+1}= \frac{N}{N+1}a_{N,m}+\gamma+h(m)
\end{align} 
where $h(m)$ is a polynomial of $(a_{n,m}-a_{n,\infty})$ for $n<N$ and $\gamma$ is some constant depending on the $a_{n,\infty}$.  The solution for (\ref{recursion}) is given by
\begin{align}
\label{h}a_{N,m} = \left(\frac{N}{N+1}\right)^m a_{N,0}+\sum_{k=1}^m\left(\frac{N}{N+1}\right)^{m-k}(\gamma+h(k)).
\end{align}
Using the induction hypothesis $h(k)$ converges exponentially to 0 as $m\rightarrow \infty$ so $a_{N,m}$ converges exponentially to $(N+1)\gamma$.  So by induction $g_n(z)$ converges exponentially to the solution for all test functions.

\section{Appendix B}
Here I would like to present some other candidate entanglement entropies for which we can determine the metric explicitly.  In each case $\tilde{S}_A = \alpha S_A$ and $2G_N\alpha/R=1.$  Note that in all cases the metric is not compatible with a vacuum gravity geometry and so it is not obvious how to reconstruct the time component.
Recall that the metric is given by
\begin{align}
ds^2 = \frac{R^2}{z^2}\left(g(z)^2dz^2+dx^2\right).
\end{align}
In all cases we us the ansatz
\begin{align}
g(z)= 1+\sum_{n=1}^\infty a_nz^n.
\end{align}
Consider a candidate entanglement entropy for a domain wall/RG flow.  
\begin{align}
\partial_\ell  \tilde{S}_A = \frac{1}{\ell}+\frac{1}{b+\ell}
\end{align}
This EE interpolates between two different radii similar to \cite{albash} equation (23), in this case $R_{IR} = 2R_{UV} (\alpha_{IR}=2\alpha_{UV})$.

Computing the first several Taylor coefficients and taking their ratio reveals that $ a_1=\nolinebreak8/b\pi$ and
\begin{align}
a_{2n+1}/a_{2n+3} = -\frac{(2n+3)b^2}{16(2n+1)}.
\end{align}
This recursion relation can be solved for $a_n$ and the resulting sum done analytically.  The result is then
\begin{align}
g(z) = 1+\frac{2}{\pi}\arctan(4z/b).
\end{align}

If we consider the case where
\begin{align}
\partial_\ell \tilde{S}_A = 1/\ell+b.
\end{align}
We can look at the table presented for general entropy and observe that
\begin{align}
\label{G}a_n = \frac{2^{n+1}\Gamma((3+n)/2)}{\sqrt{\pi}\Gamma(1+n/2)}b^n.
\end{align}
The sum coming from (\ref{G}) can be done analytically and yields
\begin{align}
g(z)=\frac{\pi+4bz\sqrt{1-4b^2z^2}+2\arcsin(2bz)}{\pi(1-4b^2z^2)^{3/2}}.
\end{align}

Next we can consider the case 
\begin{align}
\partial_\ell  \tilde{S}_A = 1/\ell+b \ell,
\end{align}

We first note that this function is odd so $g(z)$ is even.  We can look at the table and observe that
\begin{align}
&\label{H}a_{2n} = \frac{2^{4n+1}\Gamma(n+1/2)\Gamma(3/2+n)}{\pi n!(n+1)!}b^n.
\end{align}

The sum coming from (\ref{H}) can be done and results in
\begin{align}
g(z) &=  \,_2F_1(1/2,3/2;2,16bz^2).
\end{align}
Where $_2F_1$ is the hypergeometric function. 

Another case to consider is
\begin{align}
\partial_\ell  \tilde{S}_A = \frac{b}{\sin{b\ell}}.
\end{align}
One can compute the first few Taylor coefficients and notice that $a_2 =1$ and
\begin{align}
a_{2n}/a_{2(n+1)} = \left(\frac{n+1}{b(2n+1)}\right)^2.
\end{align}
This recursion relationship can be solved to obtain
\begin{align}
g(z)= \frac{2}{\pi}\EllipticK(4b^2z^2).
\end{align}
Where EllipticK is the complete elliptic integral of the first kind.  Via what amounts to Wick rotation one can obtain the expression for $\partial_\ell S_A=\beta/\sinh(\beta\ell)$ by the transformation $b\rightarrow i\beta$.

\end{document}